\documentclass[preprint,amsmath,amssymb,aps,superscriptaddress,nofootinbib,showkeys]{revtex4-2}

\usepackage[english]{babel}
\usepackage[utf8]{inputenc}
\usepackage{xr-hyper}

\usepackage[colorlinks]{hyperref}
\usepackage[T1]{fontenc}
\usepackage[dvips]{graphicx}
\usepackage{amsmath}
\usepackage{amsfonts}
\usepackage{color}
\usepackage[caption=false, justification=justified]{subfig}
\usepackage{url}
\usepackage{ulem}

\makeatletter
\newcommand*{\addFileDependency}[1]{% argument=file name and extension
  \typeout{(#1)}
  \@addtofilelist{#1}
  \IfFileExists{#1}{}{\typeout{No file #1.}}
}
\makeatother

\newcommand*{\myexternaldocument}[1]{%
    \externaldocument{#1}%
    \addFileDependency{#1.tex}%
    \addFileDependency{#1.aux}%
}

\myexternaldocument{supp_mat_rev1712}

\hypersetup{filecolor=blue}
\hypersetup{citecolor=blue}
\hypersetup{urlcolor=blue}
\hypersetup{linkcolor=blue}

\begin{document}

\title{Ultrafast spin-charge conversion at SnBi$_2$Te$_4$/Co topological insulator interfaces probed by terahertz emission spectroscopy.}
\date{\today}

\author{E.~Rongione}
\affiliation{Unité Mixte de Physique, CNRS, Thales, Université Paris-Saclay, F-91767 Palaiseau, France}
\affiliation{Laboratoire de Physique de l’Ecole Normale Supérieure, ENS, Université PSL, CNRS, Sorbonne Université, Université de Paris, F-75005 Paris, France}
\author{S.~Fragkos}
\affiliation{Institute of Nanoscience and Nanotechnology, National Center for Scientific Research “Demokritos”, G-15310 Athens, Greece}
\affiliation{Department of Mechanical Engineering, University of West Attica, G-12241 Athens, Greece}
\author{L.~Baringthon}
\affiliation{Unité Mixte de Physique, CNRS, Thales, Université Paris-Saclay, F-91767 Palaiseau, France}
\affiliation{Synchrotron SOLEIL, L’Orme des Merisiers, F-91192 Gif-sur-Yvette, France}
\author{J.~Hawecker}
\affiliation{Laboratoire de Physique de l’Ecole Normale Supérieure, ENS, Université PSL, CNRS, Sorbonne Université, Université de Paris, F-75005 Paris, France}
\author{E.~Xenogiannopoulou}
\affiliation{Institute of Nanoscience and Nanotechnology, National Center for Scientific Research “Demokritos”, G-15310 Athens, Greece}
\author{P.~Tsipas}
\affiliation{Institute of Nanoscience and Nanotechnology, National Center for Scientific Research “Demokritos”, G-15310 Athens, Greece}
\author{C.~Song}
\affiliation{Laboratoire de Physique de l’Ecole Normale Supérieure, ENS, Université PSL, CNRS, Sorbonne Université, Université de Paris, F-75005 Paris, France}
\author{M.~Mičica}
\affiliation{Laboratoire de Physique de l’Ecole Normale Supérieure, ENS, Université PSL, CNRS, Sorbonne Université, Université de Paris, F-75005 Paris, France}
\author{J.~Mangeney}
\affiliation{Laboratoire de Physique de l’Ecole Normale Supérieure, ENS, Université PSL, CNRS, Sorbonne Université, Université de Paris, F-75005 Paris, France}
\author{J.~Tignon}
\affiliation{Laboratoire de Physique de l’Ecole Normale Supérieure, ENS, Université PSL, CNRS, Sorbonne Université, Université de Paris, F-75005 Paris, France}
\author{T.~Boulier}
\affiliation{Laboratoire de Physique de l’Ecole Normale Supérieure, ENS, Université PSL, CNRS, Sorbonne Université, Université de Paris, F-75005 Paris, France}
\author{N.~Reyren}
\affiliation{Unité Mixte de Physique, CNRS, Thales, Université Paris-Saclay, F-91767 Palaiseau, France}
\author{R.~Lebrun}
\affiliation{Unité Mixte de Physique, CNRS, Thales, Université Paris-Saclay, F-91767 Palaiseau, France}
\author{J.-M.~George}
\affiliation{Unité Mixte de Physique, CNRS, Thales, Université Paris-Saclay, F-91767 Palaiseau, France}
\author{P.~Lefèvre}
\affiliation{Synchrotron SOLEIL, L’Orme des Merisiers, F-91192 Gif-sur-Yvette, France}
\author{S.~Dhillon}
\affiliation{Laboratoire de Physique de l’Ecole Normale Supérieure, ENS, Université PSL, CNRS, Sorbonne Université, Université de Paris, F-75005 Paris, France}
\affiliation{\normalfont Corresponding authors:~\href{mailto:sukhdeep.dhillon@phys.ens.fr}{sukhdeep.dhillon@phys.ens.fr},~\href{mailto:a.dimoulas@inn.demokritos.gr}{a.dimoulas@inn.demokritos.gr},~\href{mailto:henri.jaffres@cnrs-thales.fr}{henri.jaffres@cnrs-thales.fr}}
\author{A.~Dimoulas}
\affiliation{Institute of Nanoscience and Nanotechnology, National Center for Scientific Research “Demokritos”, G-15310 Athens, Greece}
\affiliation{\normalfont Corresponding authors:~\href{mailto:sukhdeep.dhillon@phys.ens.fr}{sukhdeep.dhillon@phys.ens.fr},~\href{mailto:a.dimoulas@inn.demokritos.gr}{a.dimoulas@inn.demokritos.gr},~\href{mailto:henri.jaffres@cnrs-thales.fr}{henri.jaffres@cnrs-thales.fr}}
\author{H.~Jaffrès}
\affiliation{Unité Mixte de Physique, CNRS, Thales, Université Paris-Saclay, F-91767 Palaiseau, France}
\affiliation{\normalfont Corresponding authors:~\href{mailto:sukhdeep.dhillon@phys.ens.fr}{sukhdeep.dhillon@phys.ens.fr},~\href{mailto:a.dimoulas@inn.demokritos.gr}{a.dimoulas@inn.demokritos.gr},~\href{mailto:henri.jaffres@cnrs-thales.fr}{henri.jaffres@cnrs-thales.fr}}

\newpage

\begin{abstract}
Spin-to-charge conversion (SCC) involving topological surface states (TSS) is one of the most promising routes for highly efficient spintronic devices for terahertz (THz) emission. Here, the THz generation generally occurs mainly via SCC consisting in efficient dynamical spin injection into spin-locked TSS. In this work, we demonstrate sizable THz emission from a nanometric thick topological insultator (TI)/ferromagnetic junction - SnBi$_2$Te$_4$/Co - specifically designed to avoid bulk band crossing with the TSS at the Fermi level, unlike its parent material Bi$_2$Te$_3$. THz emission time domain spectroscopy (TDS) is used to indicate the TSS contribution to the SCC by investigating the TI thickness and angular dependence of the THz emission. This work illustrates THz emission TDS as a powerful tool alongside angular resolved photoemission spectroscopy (ARPES) methods to investigate the interfacial spintronic properties of TI/ferromagnet bilayers.
\end{abstract}


\maketitle

\section{Introduction}

Spin-to-charge conversion (SCC) occurring at the interface of spin-orbit active materials through the inverse spin Hall (ISHE)~\cite{sinova2015} or the inverse Rashba-Edelstein effects (IREE)~\cite{edelstein1990,Sanchez2013} represents an important and dynamic domain~\cite{Manchon2015} owing to new potential functionalities and physical concepts. For example, strong technological interests lie in spin-orbit torque (SOT) phenomena for magnetic commutation~\cite{Mellnik2014,hellman2016,yang2018} or ultrafast terahertz (THz) wave emission~\cite{Seifert2016}. In this prospect, topological insulators~\cite{hasan2010,Manchon2015} (TIs) and their topological surface states (TSS), in particular those based on Bi compounds such as Bi$_2$Se$_3$~\cite{zhang2009}, Bi$_2$Te$_3$~\cite{Chen2009,Li2010}, SnBi$_2$Te$_4$~\cite{vergniory2015,Fragkos2021}, Bi$_{1-x}$Sb$_x$~\cite{teo2008,Hsieh2009,benia2015} or $\alpha$-Sn~\cite{rojas2016,barbedienne2019}, currently represent strong potential candidates to fulfill these applications. Indeed, such TIs or Dirac materials are known to provide very efficient SCC owing to the presence of topologically-protected surface states and spin momentum locking (SML) at their surfaces. When combined with a ferromagnetic material (FM) in bilayer structures, the IREE in TIs has recently shown magnetization reversal by SOT~\cite{Wang2017} and THz wave emission~\cite{Wang2018,Tong2021}, similar to that observed with symmetry breaking Rashba surface states~\cite{bychkov1984} in Bi/Ag systems~\cite{hoffmann2018,Qi2018}. From a fundamental point of view, the issue of a discrimination between surface IREE and bulk ISHE remains very important in that the two effects do not probe the same physical quantities, either the dynamical spin current for ISHE and spin accumulation for IREE.

Reciprocally, ultrafast time-domain techniques that detect the THz emission through the IREE (produced by spin injection and subsequent SCC)~\cite{Braun2016} have become reliable methods to probe the spin relaxation at spintronic interfaces. In this work, THz emission time domain spectroscopy (TDS) is used to generate an ultrafast spin current and subsequently a THz pulse at the interface between thin nanometric layers of Co and the TI SnBi$_2$Te$_4$, probing the spin relaxation dynamics. We show that the THz field from the TI/FM bilayer is on the same order as Co/Pt reference metallic system. We further investigated the angular dependence of the THz emission on both the magnetic (\textit{i}) and sample (\textit{ii}) in-plane orientations, showing dipolar emission features for the former (\textit{i}) experiments while an isotropic emission for the latter (\textit{ii}) is revealed as expected from an ultrafast spin-to-charge conversion process and discuss the main emission properties originating from the TI surface. Importantly the possible TSS origin of the THz emission is demonstrated by measuring the SnBi$_2$Te$_4$ thickness dependence of the emission, giving a nearly thickness independent emission, distinguishing it from any contribution from the insulating bulk.

\section{Epitaxial growth and characterization of S\lowercase{n}B\lowercase{i}$_2$T\lowercase{e}$_4$.}

\begin{figure}[!htp]
  \begin{center}
      \includegraphics[width=0.75\textwidth]{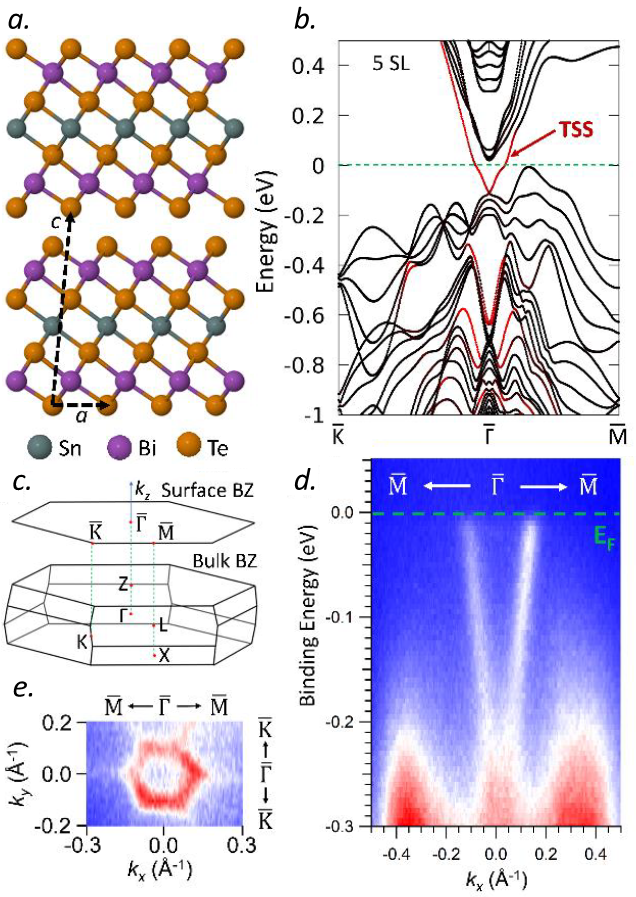}
       \caption{\textbf{TSS characterization of SnBi$_2$Te$_4$.} (a) Crystal structure of SnBi$_2$Te$_4$. Dashed arrows show the rhombohedral unit cell composed of septuple layers (SL). (b) The projected Brillouin zone (BZ) of 5~SL SnBi$_2$Te$_4$ along the $\overline{K}\overline{\Gamma}\overline{M}$ direction of the Brillouin zone. The red-colored bands are topological surface states (TSS). (c) The first BZ. (d) ARPES spectra of 5~SL SnBi$_2$Te$_4$ obtained after Te desorption for photon energy E$_{photon}=27$~eV along $\overline{M}\overline{\Gamma}\overline{M}$ and (e) the Fermi surface of the same sample.}
    \label{fig1}
  \end{center}
\end{figure}

\subsection{Material characterization}

The microstructure of SnBi$_2$Te$_4$ septuple layers (SL) was investigated by cross-sectional high-resolution scanning transmission electron spectroscopy (HR-STEM -not shown-)~\cite{Fragkos2021}. In contrast to Bi$_2$Te$_3$ quintuple layer (QL) structures, SnBi$_2$Te$_4$ data show a layered structure consisting of seven atom rows, four occupied by Te atoms and the rest by the metal atoms Bi and Sn as illustrated in Fig.~\ref{fig1}. The septuple layer structure of SnBi$_2$Te$_4$ was also confirmed by X-ray diffraction (XRD) where a pure phase of seven-layers stacking is confirmed over the entire film, indicating a tendency to form an ordered alloy where Sn ideally occupies the middle row. Details of the epitaxial growth and material characterization may be found in Suppl.~Info~\textcolor{blue}{S1} and Ref.~\cite{Fragkos2021}.

Three SnBi$_2$Te$_4$ samples were grown, whose thicknesses were determined by XRD to be 5.9, 10.2 and 16.2 nm \textit{i.e.}~5, 8 and 13~SL respectively, before being protected by a 20~nm thin amorphous Te cap. The growth process is optimized to obtain a stoichiometry allowing to pin the Fermi level with the TSS \cite{Fragkos2021} as the Fermi level position is determined by the Sn/Bi ratio. Throughout the paper, the layer thickness will be expressed between brackets in nanometers. Two different structures with the Co ferromagnetic contact were prepared, giving \textit{in fine} equivalent efficient SCC: recipe \textit{a}) where a thin Co(4) followed by a naturally oxidized AlO$_x$(3) cap layer were deposited \textit{in situ} after growth of the TI and recipe \textit{b}) in which the TI samples were transferred into another MBE chamber dedicated to angular-resolved photoemission spectroscopy (ARPES) experiments where Co is grown after the Te cap removal. In case of \textit{b}), the thickness of Co was chosen to 2~nm for SnBi$_2$Te$_4$(5.9) and Bi$_2$Te$_3$ samples. As reference samples, Co(2)/Pt(4) metallic emitters~\cite{Dang2020} were deposited by sputtering on a highly-resistive Si wafer to compare THz emission performances.

\subsection{TSS characterization}

Ab-initio calculations of the electronic band structure and the surface state dispersion were carried out using the Vienna Ab-Initio Simulation Package (VASP)~\cite{kresse1,kresse2} and projector augmented waves~\cite{blochl}. Details about the calculations are available in Suppl.~Info~\ref{S1}. SnBi$_2$Te$_4$, as its parent compound Bi$_2$Te$_3$, is known to exhibit topological nature determined by calculating the $\mathbb{Z}_2$ topological invariant in the $k_z=0$ and $k_z=\pi$ planes~\cite{wu,Fragkos2021}. This is associated to a $\Gamma_8$ and $\Gamma_9$ bands inversion between the $\Gamma$ and $Z$ points of the three-dimensional BZ resulting in a non-trivial bandgap and the appearance of surface states~\cite{Fragkos2021}. As for Bi$_2$Te$_3$, SnBi$_2$Te$_4$ crystallizes in a rhombohedral structure with space group R$\overline{3}m$ (Fig.~\ref{fig1}\textcolor{blue}{a}). Calculated projected Brillouin zone of 5~SL SnBi$_2$Te$_4$, presented in Fig.~\ref{fig1}\textcolor{blue}{b} along the $\overline{K}\overline{\Gamma}\overline{M}$ direction of the BZ (Fig.~\ref{fig1}\textcolor{blue}{c}), displays the typical TSS indicated with a red arrow. At 5 SL, we note the presence of a small indirect gap of a few meV in the calculation. However, for SnBi$_2$Te$_4$ thickness larger than 5 SL, the gap closes due to an overlap between conduction and the valence band \cite{Fragkos2021}.

\vspace{0.1in}

Specially prepared sample 5~SL SnBi$_2$Te$_4$ with a 20~nm thick Te capping was measured at the “CASSIOPEE” beamline of SOLEIL synchrotron facility after capping removal, with horizontally polarized light and a photon energy of 27~eV in order to visualize the band structure and the Fermi surface of SnBi$_2$Te$_4$. The measurements were carried out at the temperature of $15$~K and an energy resolution of 15~meV with a polar angle step of 0.2$^\circ$. Additionally to first-principles calculations, the electronic band structure of SnBi$_2$Te$_4$ along $\overline{M}\overline{\Gamma}\overline{M}$ obtained by synchrotron ARPES with photon energy of 27~eV is shown in Fig.~\ref{fig1}\textcolor{blue}{d}. We note the presence of a large gap between the position of the Fermi level and the top of the valence band of about 0.2~eV. A V-shaped band around the $\overline{\Gamma}$ point with hexagonal symmetry in $k_x-k_y$ Fermi surface (Fig.~\ref{fig1}\textcolor{blue}{e}) is clearly observed, which was previously confirmed to be a TSS, since it remained invariant to the photon excitation energy~\cite{Fragkos2021}. Note that the presence of bulk conduction band is totally absent, compared to the case of Bi$_2$Te$_3$ (Suppl.~Info~\\textcolor{blue}{S2}), where the hexagonally shaped TSS coexists with a triangularly shaped bulk band at the Fermi level, which indicates the trigonal symmetry of the material.

\section{TH\lowercase{z} emission from S\lowercase{n}B\lowercase{i}$_2$T\lowercase{e}$_4$/C\lowercase{o} and B\lowercase{i}$_2$T\lowercase{e}$_3$/C\lowercase{o}.}

\begin{figure}[!h]
  \begin{center}
    \subfloat{
      \includegraphics[width=0.45\textwidth]{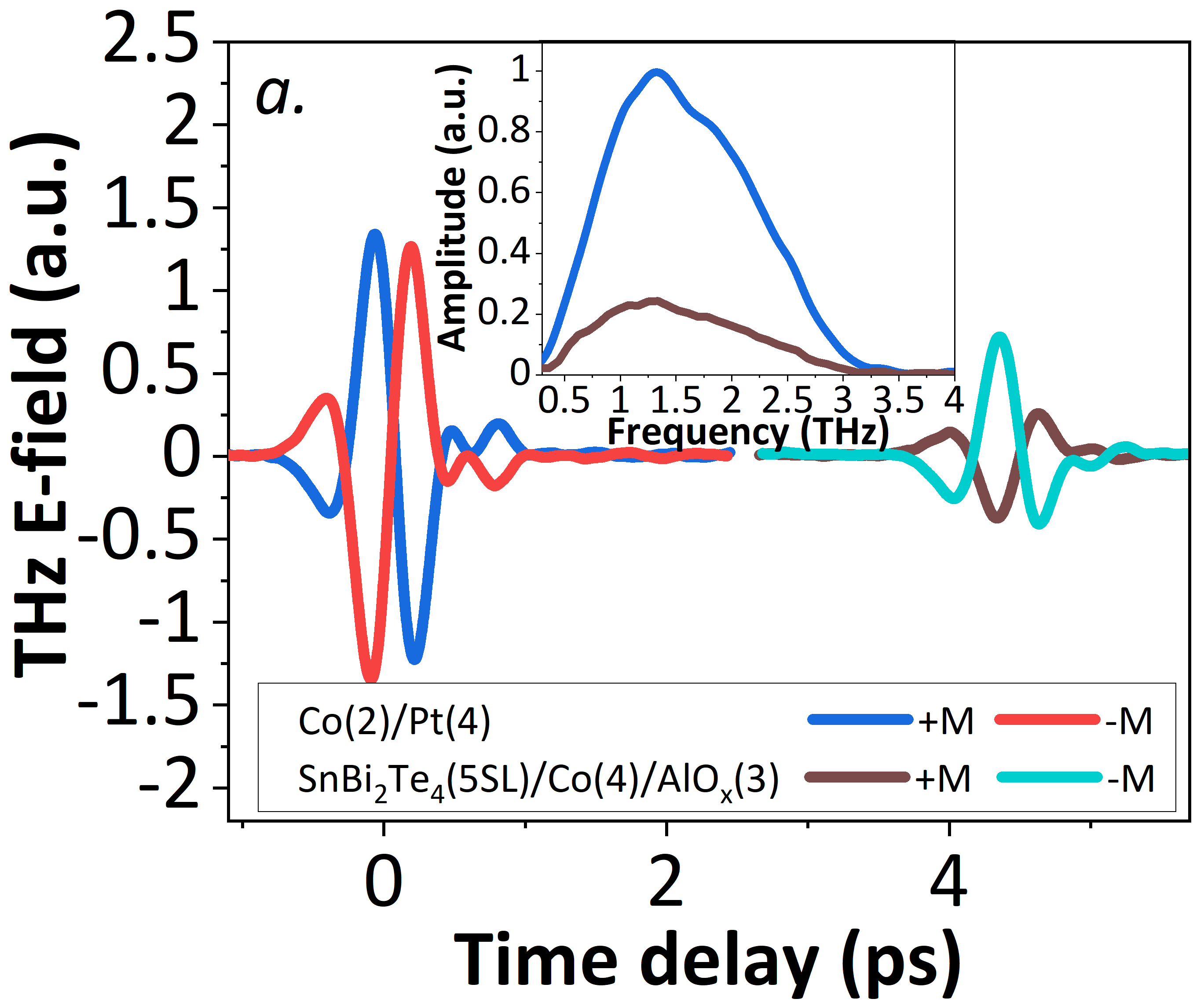}
      \label{fig2a}}
    \subfloat{
      \includegraphics[width=0.45\textwidth]{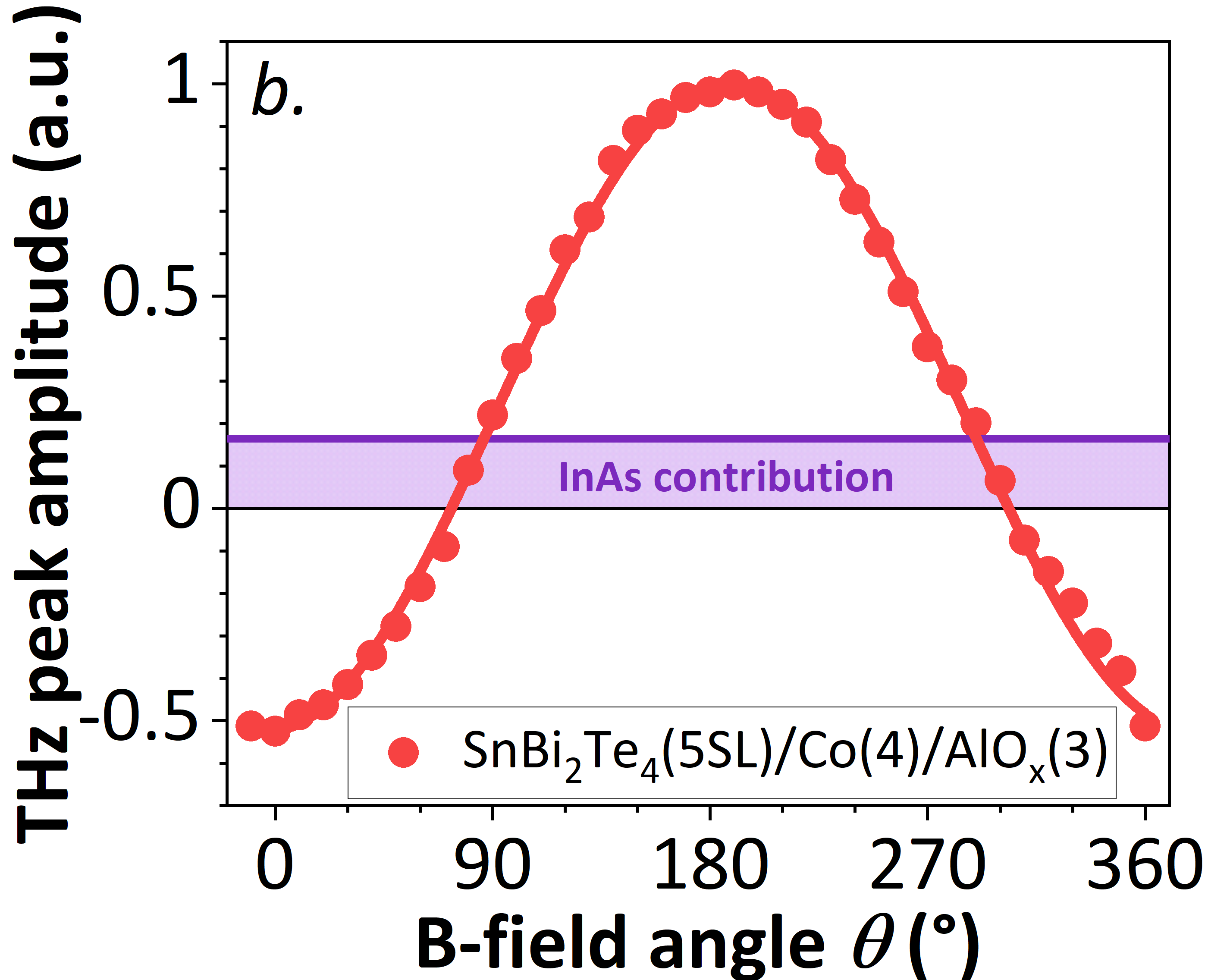}
      \label{fig2b}}
      \\
    \subfloat{
      \includegraphics[width=0.45\textwidth]{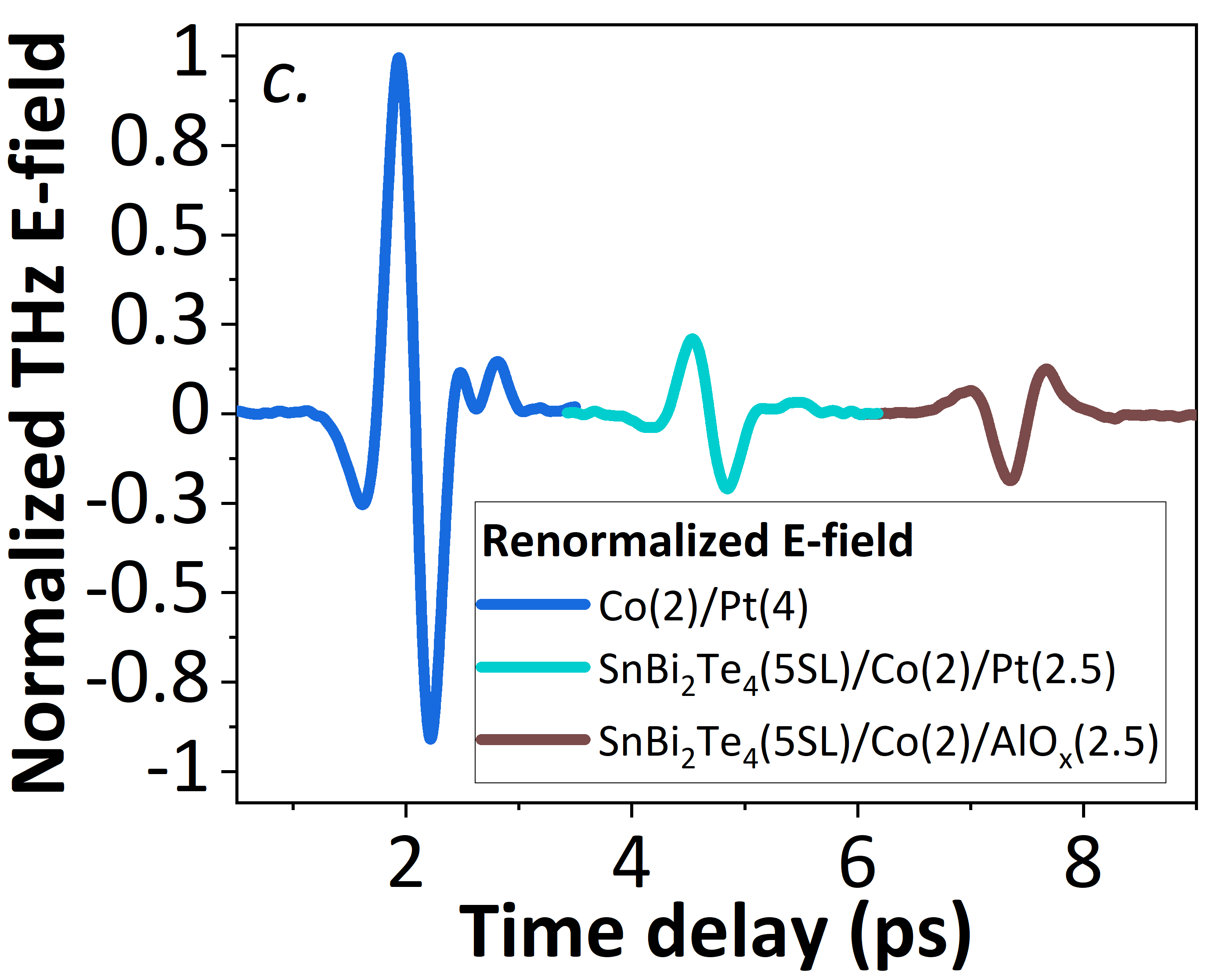}
      \label{fig2c}}
    \subfloat{
      \includegraphics[width=0.45\textwidth]{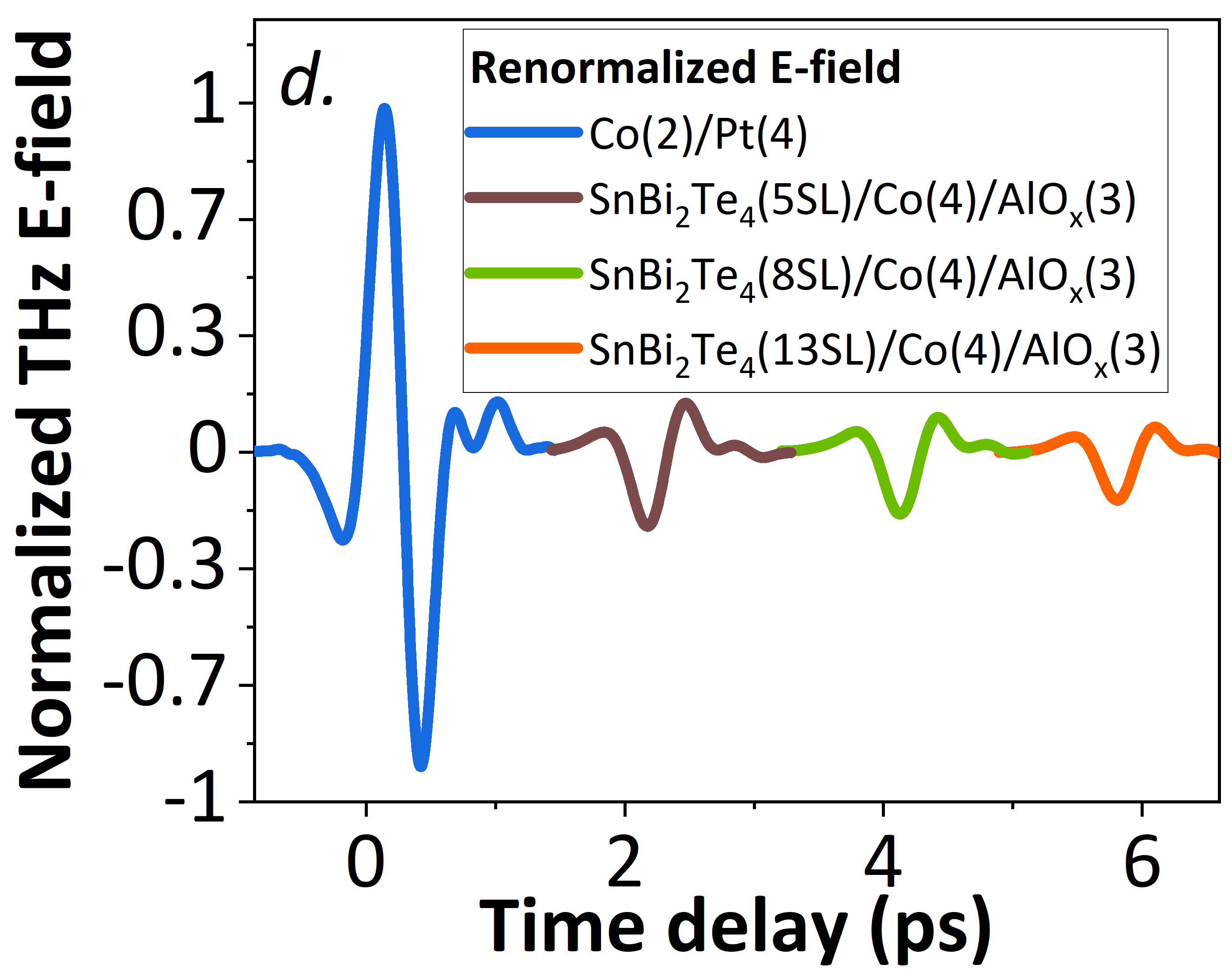}
      \label{fig2d}}
      \end{center}
    \caption{\textbf{THz emission spectroscopy acquired from SnBi$_2$Te$_4$/Co TSS mediated conversion}. (a) THz $E$-field of SnBi$_2$Te$_4$(5SL)/Co(4)/AlO$_x$(3) compared to the Co(2)/Pt(4) reference acquired under different $\pm \mathbf{M}$ polarities. The phase difference between interface states and Pt indicates identical conversion sign for SnBi$_2$Te$_4$ and Pt. The inset represents the signal in the frequency domain (Fourier transform for $+\mathbf{M}$). (b) Dipolar emission obtained by rotating in-plane magnetization $\mathbf{M}(\theta)$. The asymmetry in the peak amplitude mapping originates from magnetic and non-magnetic mixed contributions. (c) Renormalized $E$-field comparing Co/Pt and SnBi$_2$Te$_4$/Co. Note the change of THz polarization between dual SCC SnBi$_2$Te$_4$/Co/Pt and single SCC SnBi$_2$Te$_4$/Co/AlO$_x$. (d) SnBi$_2$Te$_4$ thickness dependence of the THz emission renormalized from optical radiation absorption. For (c) and (d), the electronic figure of merit has been normalized with respect to the Co(2)/Pt(4) peak amplitude. Time traces are shifted in time for clarity.}
    \label{fig2}
\end{figure}

Our investigations focus on the TSS contribution and interfacial SCC in the ultrafast SCC studied using THz-emission TDS. These experiments rely on ultrafast spin injection and their conversion into a charge current at TI material interfaces as explored recently for Bi/Bi$_2$Te$_3$ systems~\cite{Tong2021}. THz emission TDS has proven to be a relevant experimental method~\cite{Hawecker21, Gueckstock21,Cheng21} and is particularly important to probe the spin injection efficiency, the spin relaxation and related IREE mediated by interfacial states of TIs~\cite{Tong2021,Park21}. For experimental details about the THz-TDS setup, the readers may refer to Suppl.~Info~\textcolor{blue}{S1} and Ref.~\cite{Dang2020}.

\vspace{0.1in}

\begin{figure}[!htp]
\begin{center}
\subfloat{
    \includegraphics[width=0.45\textwidth]{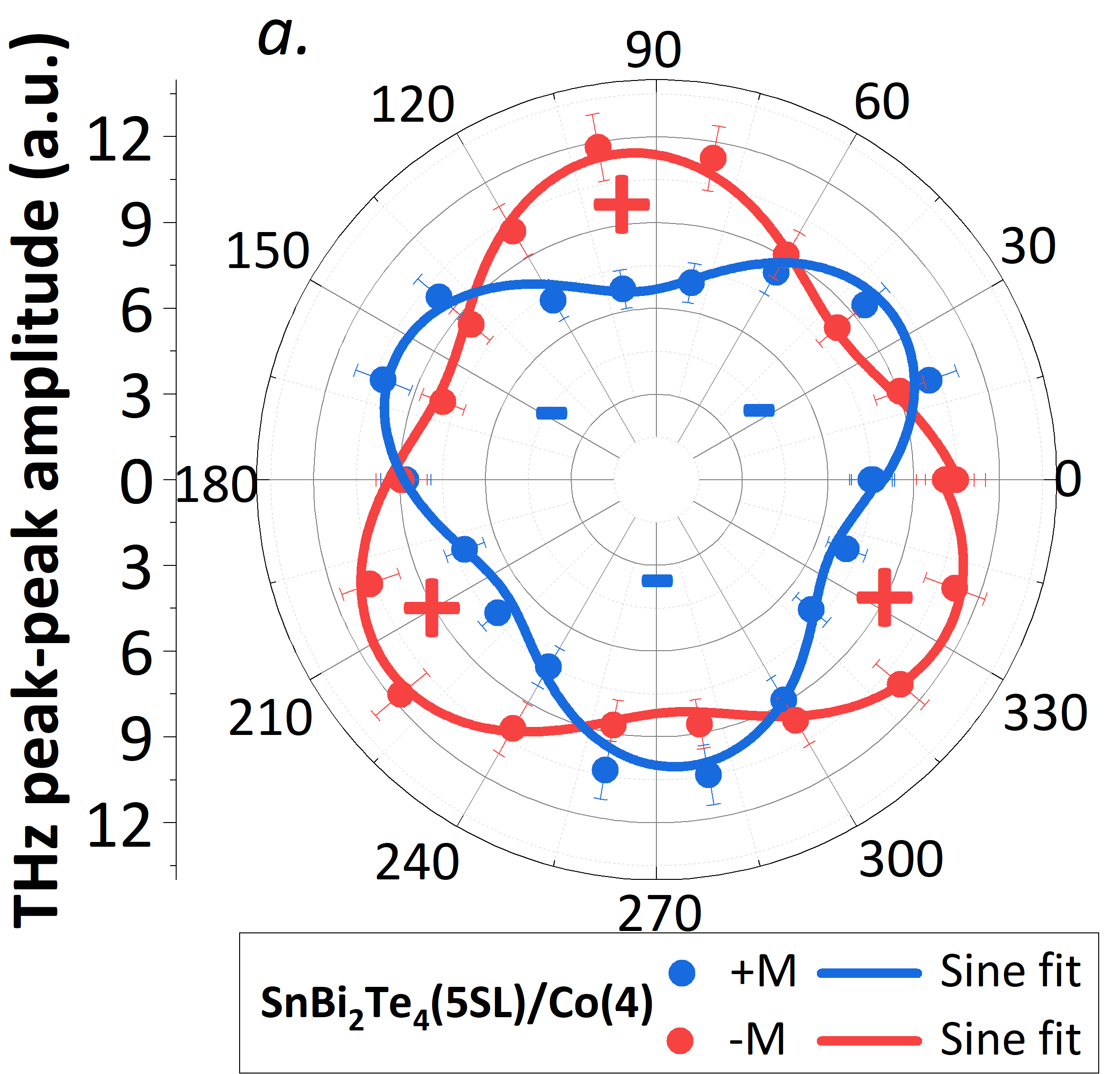}
    \label{fig3a}}
\subfloat{
     \includegraphics[width=0.445\textwidth]{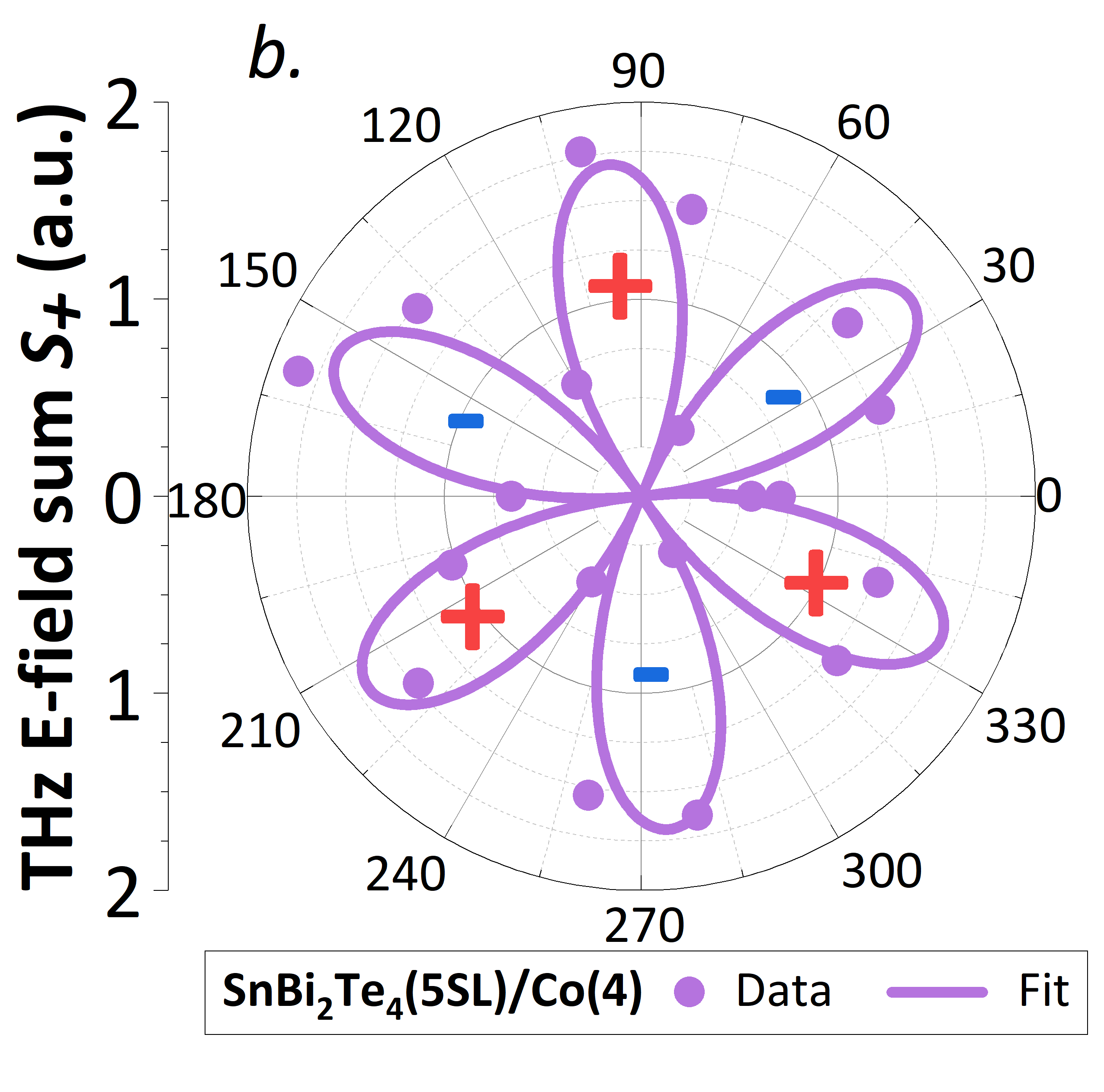}
      \label{fig3b}}
      \\
\subfloat{
     \includegraphics[width=0.45\textwidth]{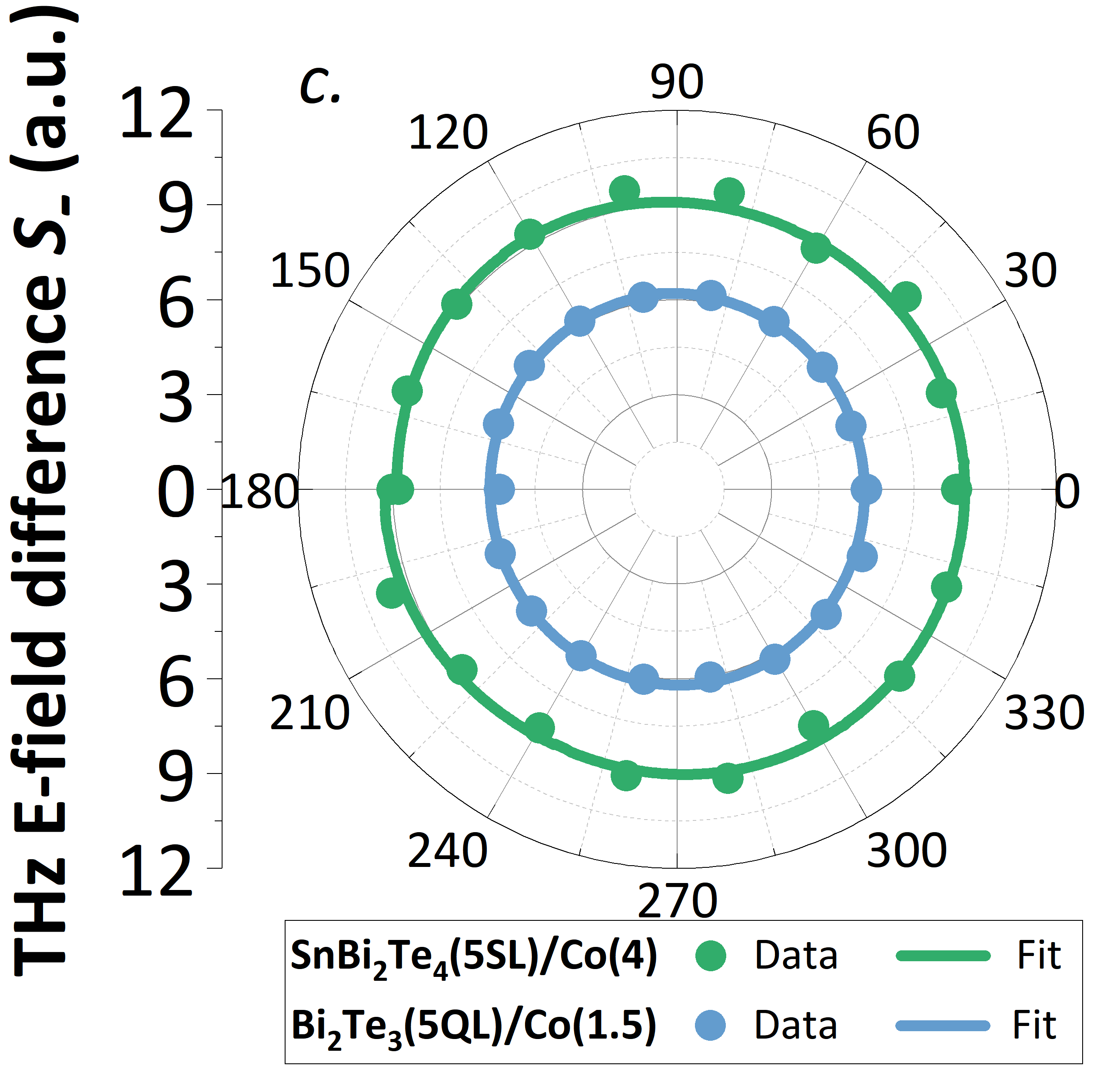}
      \label{fig3c}}
\subfloat{
     \includegraphics[width=0.46\textwidth]{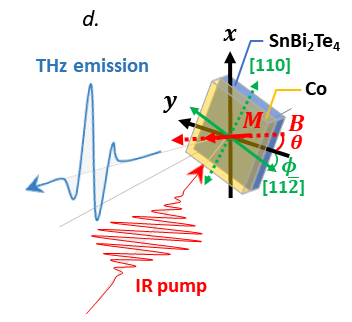}
      \label{fig3d}}
\end{center}
\caption{\textbf{SnBi$_2$Te$_4$/Co and Bi$_2$Te$_3$/Co magnetic THz emission.} (a) THz peak-to-peak amplitude \textit{vs.}~azimuthal crystalline orientation $\phi$ taken under $+\mathbf{M}$ (blue) and $-\mathbf{M}$ (red) \textit{resp.}~$\theta=0$ or $\pi$. THz even part $S_+$ (b) and odd component $S_-$
(c) giving access respectively to the non-magnetic and magnetic (renormalized) contributions in SnBi$_2$Te$_4$(5SL)/Co(4)/AlO$_x$(3) (green) and Bi$_2$Te$_3$(5QL)/Co(1.5)/AlO$_x$(2.5) (light blue). $\pm$ signs refer to the phase of the generated THz $E$-field. (d) THz-TDS setup in reflection geometry. Both IR pumping and THz collection are at normal incidence (tilted view on the scheme for clarity). The in-plane magnetization $\mathbf{M}$ orientation makes an angle $\theta$ with $e_y$. Azimuthal crystalline orientation $[11\bar{2}]$ tilts from $e_y$ with an angle $\phi$. Detection axis is along $e_x$.}
\label{fig3}
\end{figure}

This setup was used to systematically study the impact of the magnetic field orientation (about 200 mT applied in the plane by permanent magnets) and the SnBi$_2$Te$_4$ thickness on the THz emission, with the principle that $E_\text{THz}$ scales according to the fundamental quantum law of SCC, valid for both ISHE and IREE:

\begin{equation}
\mathbf{E}_{\text{THz}} \propto \frac{\partial \mathcal{J}_c}{\partial t} \propto \alpha_{\text{SCC}} ~\frac{\partial}{\partial t} \left(\mathcal{J}_s \times \mathbf{M} \right)
\label{eq1}
\end{equation}
where $\mathcal{J}_c$ and $\mathcal{J}_s$ are respectively the in-plane charge and normal spin current and $\mathbf{M}$ the magnetization vector. For ISHE, the conversion figure of merit is the so-called spin Hall angle $\alpha_{\text{SCC}}=\theta_{\text{SHE}}$. In TIs, one expects that the interfacial conversion to occur by IREE on the related SML Fermi surfaces of the TSS on the characteristic inverse Edelstein length $\alpha_{\text{SCC}} = \lambda_{\text{IREE}}$~\cite{rojas2016,Fert19}. In the following, we define two angles referenced with respect to $e_y$ axis: $\theta$ the direction of the in-plane magnetization $\mathbf{M}(\theta)$ (by rotating the applied magnetic field $B$) and $\phi$ tilt of the $[11\bar{2}]$ azimuthal crystallographic direction (rotating the sample) as illustrated in Fig.~\ref{fig3d}. Detection axis is along $e_x$.

\subsection{THz emission from SCC in SnBi$_2$Te$_4$.}

Fig.~\ref{fig2} displays the THz $E$-field acquired from the different SnBi$_2$Te$_4$/Co samples compared to our Co(2)/Pt(4) reference~\cite{Dang2020} (offset in time for clarity). The sampled $E_{\text{THz}}$ obtained for SnBi$_2$Te$_4$ at $\pm \mathbf{M}$ polarities gives rise to an opposite THz polarization as expected for SCC (Eq.~\ref{eq1}). More interestingly, the THz $E$-field phase is flipped by $\pi$ while comparing THz pulse from Co/Pt and SnBi$_2$Te$_4$/Co with the same $\mathbf{M}$ orientation. By noticing the IR pump hits first the Pt layer then Co in the first sample and conversely hits first AlO$_x$ then Co and SnBi$_2$Te$_4$ in the latter, it indicates a same SCC sign for SnBi$_2$Te$_4$ and Pt owing to the opposite spin flow from Co \cite{Wang2018,Chen2021}. Moreover, from the maximal THz amplitude and from its Fourier transform, one notices that SnBi$_2$Te$_4$/Co $E_{\text{THz}}$ is about 25\% of the Co/Pt reference emitter. Fig.~\ref{fig2b} shows the signature of the dipolar THz emission from SnBi$_2$Te$_4$(5SL)/Co(4)/AlO$_x$(3) sample, by varying $\mathbf{M}(\theta)$. One observes a quasi perfect sine shape dependence of $E_{\text{THz}}$ \textit{vs.} $\theta$, representative of the SCC process; and not from other spurious optical rectification effects on the surface crystal (\textit{e.g.} second harmonic $\chi^{(2)}$). Indeed, such a sine angular dependence is generally a signature of ISHE~\cite{Seifert2016,Dang2020} and has also recently been observed in the case of Rashba conversion (IREE) in Bi/Ag~\cite{hoffmann2018,Qi2018} and in TSS~\cite{Wang2017,Tong2021}, thus questioning about the exact SCC origin of SnBi$_2$Te$_4$ THz emission. Moreover, from experiments, one can notice the presence of a small vertical shift of the THz signal from the zero angular baseline which may be explained by an additional non-magnetic contribution from the 35~nm-thick InAs buffer~\cite{reid2005} (additional discussions are available in Suppl.~Info~\textcolor{blue}{S3}).

In the next, we will isolate the magnetic THz signal to express and quantify the SCC. Fig.~\ref{fig2c} shows the emitted electric field ($E_{\text{THz}}$) of the SnBi$_2$Te$_4$(5SL)/Co(2)/Pt(2.5) where a 2.5~nm thick THz active Pt layer is grown as an overlayer instead of 2~nm oxidized Al (AlO$_x$) used to prevent oxidation. The Co thickness has been reduced down to 2~nm to limit THz absorption. In this case, one notices the strong reproducibility of TI-based THz emission between two different growth recipes. Indeed, the second series of samples (recipe \textit{b}) investigated here have underwent Te capping desorption, while still keeping their TSS and their SML properties. $E_{\text{THz}}$ displayed on Fig.~\ref{fig2c} has been renormalized for both THz and infrared (IR) absorption, in order to fairly compare their \textit{electronic} SCC figure of merit. For this goal, we have estimated the bulk conductivity of SnBi$_2$Te$_4$ in the vicinity of $2.35 \times 10^{5} ~\Omega^{-1}.\text{m}^{-1}$ \textit{via} Hall effect measurements~\cite{Fragkos2021} (more details on the procedure are available in Suppl.~Info~\textcolor{blue}{S5}). $E_{\text{THz}}$ are about equal in peak-to-peak amplitude revealing two specific features compared to SnBi$_2$Te$_4$(5SL)/Co(4)/AlO$_x$(3): \textit{i}) the THz polarization is reversed, showing now a predominant signal from ISHE of Pt (dual SCC occurs here with both ISHE in Pt and expected interfacial SCC for SnBi$_2$Te$_4$/Co) and \textit{ii}) renormalized $E_{\text{THz}}$ emission of SnBi$_2$Te$_4$(5SL)/Co(2) is estimated around 20\% of the one of Co/Pt. Such THz polarization inversion confirms again that the SCC sign of SnBi$_2$Te$_4$ is positive as the one of Pt. Note that in this simple descriptive scheme, the pure InAs contribution cancels out from the difference of the two THz signals.

Fig.~\ref{fig2d} presents the THz signal thickness dependence of SnBi$_2$Te$_4$($x$) renormalized from both optical and THz absorption. Interestingly, $E_{\text{THz}}$ from SnBi$_2$Te$_4$($x$)/Co(4) is about $20$\%, $15$\% and $10$\% ($\pm 2$\%) of the Co(2)/Pt(4) SCC (positive SCC sign), respectively for 5~SL, 8~SL and 13~SL. As well as demonstrating the high reproducibility of the sample quality (in this study, samples from recipe \textit{a} were measured) and SCC conversion for this TI, the thickness dependence also permits to determine the process of this SCC. Indeed, THz emission that would depend only on ISHE would result to a strong thickness dependence associated to the establishment of the SCC over a length scale equaling the characteristic hot electron relaxation length which is not the case here. On the other hand, SCC owing from IREE would preferably lead to a constant renormalized THz signal as far as the TSS are formed in the thickness of the material. This seems to be the case here where in the case of pure Bi, for example, the formation of such TSS is known to occur over a very short distance near the $\Gamma$ point~\cite{Ishida_2016}. The slight difference observed in the signal amplitude in the thickness series may traduce: \textit{i}) a certain SnBi$_2$Te$_4$ conductivity variation during the deposition leading to a differential THz absorption, \textit{ii}) TSS characterized by a different Fermi surface warping~\cite{fu2009} (Suppl.~Info~\textcolor{blue}{S6}), or \textit{iii}) a reduced SHE bulk contribution to the SCC originating from the valence bulk states nearby the Fermi level. However, this last scenario has to be discarded in the case of a gapped semiconductor as considered here as an experimental gap of 0.2 eV is observed by ARPES (Fig.~\ref{fig1}\textcolor{blue}{d}) on the bare SnBi$_2$Te$_4$. Concerning the samples investigated here, the reduction of the SnBi$_2$Te$_4$ thickness would lead to an increase of the gap by quantization effects accompanied with a strong reduction of the spin current injected from Co. This cannot explain a THz signal as large as a fraction of the one of metallic Co/Pt as observed. A direct proof of a pure TSS contribution in THz emission would be the demonstration of a THz signal in the limit of the typical extension of such TSS in the film thickness. However, previous work performed on the same material family (Bi$_2$Se$_3$) has shown that the TSS persist down to 3~QL~\cite{tsipas2014} which make almost impossible, in our case, this definite proof on thinner samples from this series.

\subsection{Crystallographic symmetries of TI/FM SCC THz emission.}

To further highlight the TI interface origin of the THz emission, we now focus on the specific SCC symmetries with respect to the crystallographic azimuthal orientation $\phi$, owing to the specific topology of the Dirac cone~\cite{fu2009,Vajna2012,Johansson2016}. Indeed, from the linear response theory~\cite{litvinov2020,marder2000}, one can show that the SCC mediated by IREE on a (111)-TSS possessing a hexagonal structure of the Fermi contour should lead to an isotropic emission with respect to $\phi$, which amplitude depends on the Fermi surface warping (Suppl.~Info~\textcolor{blue}{S6}). To access experimentally this dependence, we performed by varying the crystalline orientation angle $\phi$ while keeping $\theta=0$ or $\pi$ \textit{i.e} configurations resulting in the maximum THz emission as illustrated in Fig.~\ref{fig3d}: doing so, we would like to address THz crystalline TSS emission symmetries.

Experimental and analysis procedures were used to extract, from the SnBi$_2$Te$_4$(5SL)/Co(4)/AlO$_x$(3) sample, the magnetic and non-magnetic contributions to the overall THz emission (refer to Suppl.~Info~\textcolor{blue}{S7}). Fig.~\ref{fig3a} displays the subsequent angular dependence of $E_{\text{THz}}$ for the two opposite magnetizations $\pm \mathbf{M}$ polarities. Unlike the dipolar emission $\mathbf{M}(\theta)$, one observes a clear emission anisotropy keeping the same THz phase $\Phi=\Phi_{\text{THz}}$ within the range $\phi=[0-2\pi]$ in each of the two configurations~\textit{i.e.} positive in red for $+\mathbf{M}$ (\textit{resp.} negative in blue for $-\mathbf{M}$). Such anisotropy is characterized in each case by an amplitude of about 15 to 20\% of the averaged emission (note the scale of the $y$-axis in the polar plot). The maximum of the signal for $+\mathbf{M}$ corresponds to a minimum of $-\mathbf{M}$ and \textit{vice-versa}. This reflects a additional non-magnetic contribution with a $2\pi/3$ angle periodicity.

We note $\Phi_{\pm \mathbf{M}}$ as the THz phase acquired for $\theta=0$ ($+\mathbf{M}$) and $\theta=\pi$ ($-\mathbf{M}$) as illustrated in the plots as $\pm$ notation. By taking the THz signal sum (even) $S_+=(\Phi_{+\mathbf{M}} S_{+\mathbf{M}} + \Phi_{-\mathbf{M}} S_{-\mathbf{M}})/2$ and  difference (odd) $S_-=(\Phi_{+\mathbf{M}} S_{+\mathbf{M}} - \Phi_{-\mathbf{M}} S_{-\mathbf{M}})/2$, we can access, respectively, the non-magnetic and magnetic (spin injection) contributions. The results are reported respectively in Fig.~\ref{fig3b} for even $S_+$ and Fig.~\ref{fig3c} for odd $S_-$. Moreover, one can totally exclude the Co self-emission due to interfacial spin-flip like observed in other equivalent systems: such contribution is generally limited to a factor of 1/20 compared to the Co/Pt reference~\cite{wu2019} (see Suppl.~Info~\ref{S9}). It becomes clear that the spin injection contribution displays an almost perfect isotropic character (Fig.~\ref{fig3c}) that we attribute to SnBi$_2$Te$_4$/Co interfacial SCC mediated at the interface with the TI. Fig.~\ref{fig3c} also presents the magnetic (spin injection) contribution of $E_{\text{THz}}$ acquired on the companion Bi$_2$Te$_3$(5QL)/Co(1.5) sample in the same experimental conditions and analysis procedure. Compared to SnBi$_2$Te$_4$(5SL), the THz magnetic emission is found to be about 30\% lower. This can be explained by the merging of the TSS with the bulk bands crossing the Fermi level for this particular Bi$_2$Te$_3$ stoichiometry; or possibly by a different warping of the hexagonal Fermi surface.

Finally, we discuss the non-magnetic contribution given by the even component $S_+$ showing a six-fold emission symmetry (Fig.~\ref{fig3b}). This is ascribed to optical rectification effects at the InAs(111) interface from the second order non-linearity $\chi^{(2)}$ term~\cite{reid2005}. This conclusion is also supported by the angular signature of the THz data from the reference InAs(111) buffer (Suppl.~Info~\textcolor{blue}{S4}). We identify the mean non-zero value component to a possible photo-Dember effect that may occur from a slight misorientation in the experimental configuration reflection geometry (Suppl.~Info~\textcolor{blue}{S8}).

\section{Conclusions}

To conclude, we have demonstrated THz emission from a series of MBE grown SnBi$_2$Te$_4$/Co structures whose stoichiometry of SnBi$_2$Te$_4$ is carefully designed in order to isolate the TSS from the bulk bands, in contrast to previous studies on TIs. Using THz emission TDS, we clearly show a sizable ultrafast spin injection at the TI/Co interfaces with a SCC efficiency of the same order as the Co/Pt metallic reference system and of the same sign. The THz emission displays an isotropic emission character with respect to the crystallographic orientation as expected from the Dirac-like properties of TSS (IREE). Nearly constant THz emission amplitude is observed when varying the SnBi$_2$Te$_4$ thickness in a sample series, indicating a rather short evanescent length, typically shorter than 5 nm. Comparisons between SnBi$_2$Te$_4$/Co and Bi$_2$Te$_3$/Co samples show an enhanced emission from the former, which may be explained by an ultrafast SCC at the interfaces of TI/FM and their TSS in the case of SnBi$_2$Te$_4$ of selected stoichiometry. We cannot however totally discriminate a small contribution from ISHE originating from more delocalized valence band states in the layer thickness. THz emission spectroscopy reveals to be one of the most valued method to probe the sub-picosecond dynamics of SCC in complex systems composed of advanced quantum surfaces and ferromagnetic contact. The presence of InAs substrate prevents us to possible to put in evidence the warping of the Fermi surface.

\section*{Acknowledgments}

The authors thank T.~H.~Dang for her help in the data analysis. The authors thank G.~Bierhance and T.~Kampfrath for fruitful discussions. We acknowledge the Horizon 2020 FETPROAC Project No.~SKYTOP-824123 “Skyrmion—Topological Insulator and Weyl Semimetal Technology”. We acknowledge financial support from the Horizon 2020 Framework Programme of the European Commission under FET-Open grant agreement No.~863155 (s-Nebula).

\bibliographystyle{ieeetr}
\bibliography{biblio}

\end{document}